\documentclass[aps,prl,reprint,groupedaddress]{revtex4-1}
\usepackage{graphicx}
\usepackage{amsmath}

\begin{document}

\title{Eigenvalues of the Homogeneous Finite Linear One Step Master Equation: Applications to Downhill Folding}
\author{Thomas J. Lane}
\affiliation{Department of Chemistry, Stanford University}
\author{Vijay S. Pande}
\affiliation{Department of Chemistry, Department of Computer Science, and Biophysics Program, Stanford University}
\date{\today}

\begin{abstract}
Motivated by claims about the nature of the observed timescales in protein systems said to fold ``downhill'', we have studied the finite, linear master equation
\[
\dot{p_n} = k_f p_{n-1} + k_u p_{n+1} - (k_f + k_u) p_n
\]
which is a model of the downhill process. By solving for the system eigenvalues, we prove the often stated claim that in situations where there is no free energy barrier, a transition between single and multi-exponential kinetics occurs at sufficient bias (towards the native state). Consequences for protein folding, especially the downhill folding scenario, are briefly discussed.
\end{abstract}


\maketitle

\section*{Introduction}

In studies of protein folding, it is often claimed that at sufficient native bias we can expect the system kinetics to become multi-exponential or ``downhill" \cite{Bicout:2000vu, Bryngelson:1995hq, Eaton:1999wo}. While a relatively simple statement, it is often presented without proof nor even heuristic justification.

There are many reasons why such a statement is not accompanied by evidence. Foremost is the fact that protein folding is often modeled by one dimensional reaction coordinates, on which dynamics are governed by a Smoluchowski equation. Even in situations where such a model is appropriate, in the continuous limit it is very hard to determine the system's timescale spectrum. What one is truly interested in are the eigenvalues of the master equation, which is the Smoluchowski representation are blurred into a continuous spectral density by the limiting procedure used to transform a discrete master equation into a continuous propagator \cite{vanKampen:2007vs}.

This blurring is often mirrored in experimental studies, where kinetic traces, $A(t)$, have sometimes been fit to stretched exponentials
\begin{equation}\label{str-exp}
A(t) = A_0 e^{- \alpha t^\beta}
\end{equation}
while such a model has few free parameters ($\alpha$, $\beta$), the underlying phenomena leading to such a kinetic response is typically complex (see \emph{e.g}. the classic work \cite{PALMER:1984ud}). If the underlying physics of the system is stationary and Markovian, then we expect any kinetic response to be composed of elementary processes that decay as exponentials
\begin{equation}\label{sum-exp}
A(t) = \sum_i A_i e^{ - \alpha_i t}
\end{equation}
One can readily verify by Taylor expansion that such a series can reproduce any monotonically decreasing function given a suitable choice of positive $\alpha_i$ and $A_i$. Specifically, if there are many closely spaced exponentials a stretched exponential form can arise. The model presented here explains one mechanism by which a stretched exponential could emerge.

It should be noted that in protein folding, most stretched exponentials are equally well fit by a summation of two or more elementary exponentials \cite{Sabelko:1999wz, Voelz:2011iz}, and thus researchers have a choice for how to fit their data. While the stretched exponential form may be more convenient for fitting data, it is phenomenological; the summation of exponentials is microscopic. In some sense, then, the stretched exponential form is less appealing, because it leaves out a connection to the microscopic physics describing the system. This is why models like the one presented here are useful -- they represent an intermediate step in connecting phenomenology and theory.
%
\begin{figure}
\includegraphics[width=8cm]{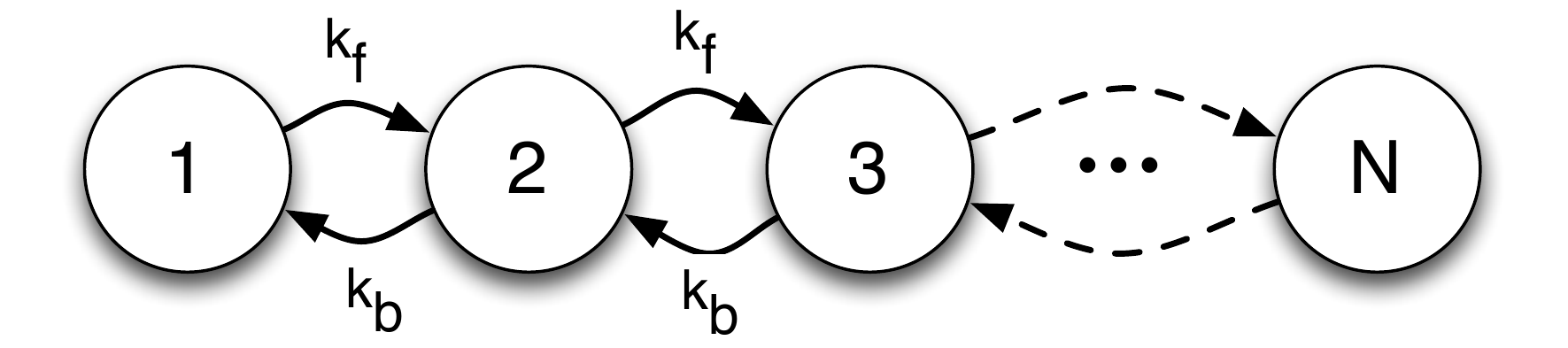}
\caption{Illustration of the one-step process under consideration. The model consists of a linear sequence of $N$ states, with a forward rate $k_f$ and backward rate $k_b$.}
\label{diagram}
\end{figure}

Since multi-exponential kinetics have been observed in a number of experimental studies of protein folding \cite{Sabelko:1999wz, Liu:2007ij, Sadqi:2006ia}, understanding the origins of this behavior is an imperative goal for theory. This is especially true since the experimental community has argued over what experimental observations are sufficient to claim downhill behavior \cite{Zhou:2007cd, Cho:2008vy, Huang:2009vp}. Moreover, downhill behavior is considered a key prediction of theory, though the original theories describing protein folding were ambiguous as to whether or not a large free-energy bias resulted in multi-exponential kinetics \cite{Bryngelson:1995hq}.

Some authors working in folding have been able to work around this difficulty, e.g. Bicout and Szabo showed that there were discrete exponential timescales in the mean first passage time distribution to the native state \cite{Bicout:2000vu}. Such a treatment, however, cannot account for ergodicity in a systematic manner and therefore suffers from some limitations.

Here we present proof that there is a transition from single to multi-exponential kinetics as a system moves from no bias to a heavy bias towards a native state. We do this by calculating the eigenvalues of a finite, homogeneous, linear one-step process
\[
\dot{p_n} = k_f p_{n-1} + k_u p_{n+1} - (k_f + k_u) p_n
\]
illustrated in Figure \ref{diagram}. In this model, we consider the transitions of an ensemble of walkers (proteins) between a series of $N$ states over time. In each state, the walkers move to ``forward'' (more native) at a rate of $k_f$, and backwards (less native) with rate $k_b$. Our central result is a closed-form expression for the eigenvalues of the master equation described, which give the rates of relaxation that would be measured in a bulk experiment performed on such a system.

\begin{figure}
\includegraphics[width=8cm]{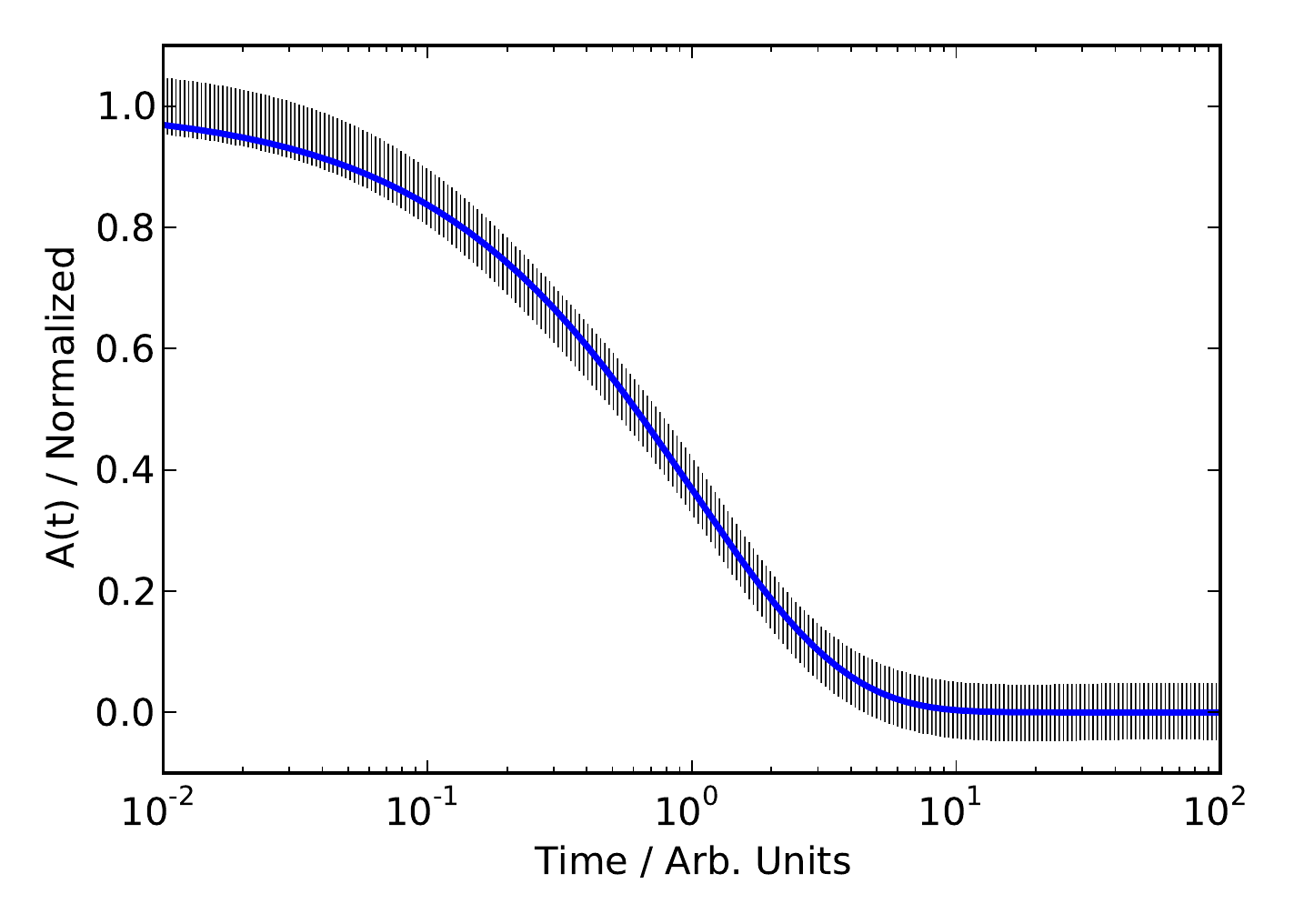} 
\caption{The one-step model's relaxation spectrum can account for a stretched exponential. Plotted in blue is a stretched exponential (\ref{str-exp}) with $\alpha = 1$, $\beta = 0.75$ and in black vertical lines, the weighted sum of exponential decays (\ref{sum-exp}), with $\alpha_i$ from the eigenvalues of the one-step model. Here $k_f / k_u = 1.4$ and $N=100$. Weights $A_i$ were chosen to match the stretched exponential.}
\label{stretched}
\end{figure}

This process is a discrete analog to the one-dimensional reaction coordinate picture of protein folding, where in the downhill scenario the free energy decreases monotonically as one approaches the native state. The presented model has a discrete timescale spectrum, as observed experimentally in studies of protein folding. Further, the discrete spectrum is fine enough to describe a stretched exponential (Fig.~\ref{stretched}). By scaling the the folding bias, we show that there is a clear transition from single-exponential kinetics at low biases (flat/golf-course profiles) to multi-exponential kinetics at large biases.

Here, we ignore the larger (and more interesting) questions of whether or not a linear model such as the one presented is a good model of folding. Instead, we wish to simply show that a sufficient native bias does lead to multi-exponential kinetics, while in lieu of any bias one obtains single exponential behavior. We hope that the analysis presented here will help clean up some of the confusion for under what conditions we can expect such behavior.

As a final note, while the model presented is quite simple, and should be common in fields other than protein folding we have not been able to find a clear solution presented in the literature. We found this quite surprising considering this model may have wide applicability. We present the diagonalization of an important class of tridiagonal matrices that we expect appears in many fields. Note that solutions for similar models have been published, but not this exact one \cite{Yueh:2005ux, Kouachi:2008vy}. Hopefully the solution presented here can be used in other applications.

First, we present the mathematical description of the model and its solution. We proceed to discuss interesting aspects of this solution. The reader uninterested in mathematical detail can stop there - the rest of the paper is devoted to a proof of the solution.

\begin{figure}
\includegraphics[width=8cm]{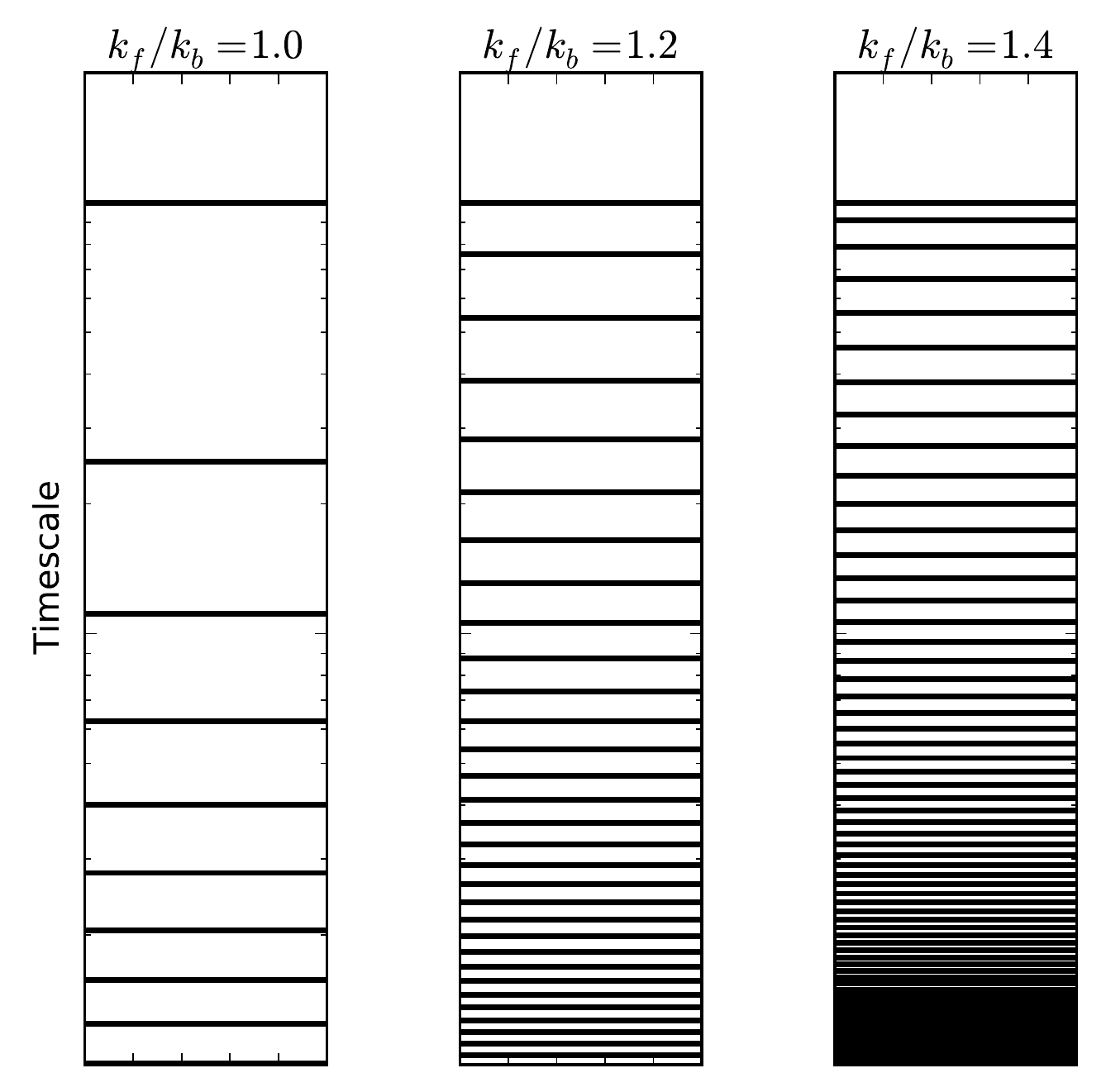}
\caption{Illustration of the analogy between the model considered and ``downhill'' free energy profiles. Top: The timescale spectra at various levels of bias, showing the transition from a large timescale separation at low bias to a more finely space spectrum. Bottom: The free energy profiles that correspond to these spectra, labels are the value of bias $k_f / k_b$.}
\label{spectra-profiles}
\end{figure}

\section{Solution and Analysis of the Model}

The one-step model just described is equivalent to the master equation
\[
\frac{dP}{dt} = KP
\]
where $P$ is an $N$-vector of the state populations, and $K$ is a rate matrix of the form
\[
K = \left[ \begin{array}{ccccc}
-k_f & k_f & 0 & \cdots & 0\\
k_b & -(k_f + k_b) & k_f  &  & \vdots\\
\vdots &  &  \ddots  & & \vdots\\
\vdots & & k_b & -(k_f + k_b) & k_f  \\
0 & \cdots & 0 & k_b & -k_b
\end{array} \right]
\]

In what follows, we will show that the eigenvalues of this matrix are
\begin{eqnarray} \label{spectrum}
\lambda_{k} =
\begin{cases}
0 & k = 1 \ \ \\
 -(k_f+k_b) + 2 \sqrt{k_f k_b} \cos \left( \frac{k \pi}{N} \right) & k = 2,...,N \ \
\end{cases}
\end{eqnarray}
where $N$ is the size of $K$, and $k$ indexes the eigenvalues. This formula shows the eigenvalues of such a matrix are equally spaced at intervals of $2 \sqrt{k_f k_b}$.

Using this equation, we can easily calculate the conditions for single versus multi-exponential behavior. These eigenvalues $\lambda_k$ are rates of relaxation, and are directly related to the characteristic relaxation timescales of the system
\[
\tau_k = - \frac{1}{\lambda_k}
\]
looking at the spacing of these timescales tells us whether or not the system would appear single or multi-exponential to an experiment (Fig.~\ref{spectra-profiles}). If there is a large gap between the first ($\tau_2$) and second ($\tau_3$) timescales, then the system is ``single'' exponential, while if the gap is small the system is multi-exponential. Note that $\tau_1 = \infty$, and represents the stationary solution, such that there will always be $N-1$ observable timescales, regardless of whether we talk about the dynamics being single or multi-exponential.

\begin{figure}
\includegraphics[width=8cm]{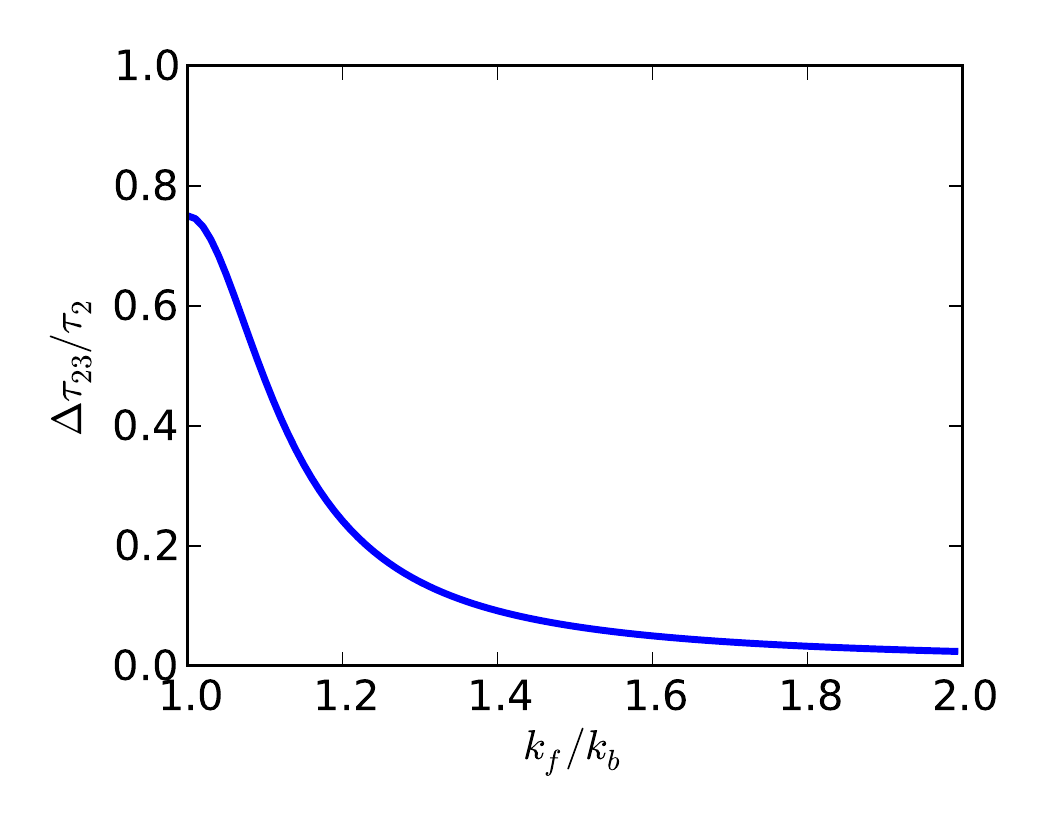}
\caption{The kinetic isolation $\Delta \tau_{23} / \tau_2 $ of the model as a function of the ratio of the forward to backward rate, $k_f / k_b$. The isolation is normalized by the longest timescale $\tau_2$. There is a clear transition from single to multi-exponential behavior at around $k_f / k_b \approx 1.2$}
\label{spectral-gap}
\end{figure}

To help quantify the two regimes of interest, we introduce the \emph{kinetic isolation}, $\Delta \tau_{23} / \tau_2$, where $\Delta \tau_{23} \equiv \tau_2 - \tau_3$. This value provides a normalized measure between $0$ and $1$ of the size of the separation between the first and second observable timescales. With Eq.~(\ref{spectrum}) in hand, the calculation of the kinetic isolation is trivial - Figure \ref{spectral-gap} shows the kinetic isolation as a function of increasing bias. The figure clearly shows that around $k_f / k_b \approx 1.2$, a relatively small bias, there is a dramatic shift from single exponential behavior (large kinetic isolation) to multi-exponential behavior (small isolation).

This figure represents our central result, which is simply verification of the claim that as a one-dimensional model moves from a ``golf-course'', or unbiased free energy landscape, the kinetics shift from single to multi-exponential.

\section{Solution of the Model}

What follows is simply the mathematical justification for Equation \ref{spectrum}, the closed-form expression for the eigenvalues of $K$.

Laplace's expansion in minors gives the characteristic polynomial of a matrix recursively in terms of the minors of the matrix. For generic tridiagonal matrix $A = (a_{i,j})$, it can be shown that the the expansion for the characteristic equation takes the simplified form
\begin{equation} \label{continuant}
|A_n| = a_{n,n} | A_{n-1} | - a_{n-1,n}a_{n,n-1} | A_{n-2} |
\end{equation}
with
\[
 | A_0 | = 1 \ \ \mathrm{and} \ \  | A_1 | = a_{1,1}
\]
where $| \cdot |$ denotes a matrix determinant and $A_i$ indicates the $i^{\mathrm{th}}$ principle minor of $A$, that is the matrix formed by taking the first $i$ rows and columns of $A$.

We begin by re-writing $K$ for notational convenience. Let
\[
A = \left[ \begin{array}{ccccc}
-a & a & 0 & \cdots & 0\\
b & -(a + b) & a  & & \vdots\\
\vdots &  &  \ddots  & & \vdots\\
\vdots & & b & -(a + b) & a  \\
0 & \cdots & 0 & b & -b
\end{array} \right]
\]
with $a= \sqrt{k_f / k_u }$ and $b=\sqrt{k_u / k_f }$, for reasons that will be clear in a moment. By this definition, $K= (k_f k_u)^{-1/2}A$, and also $ab = 1$, which will be a key later.

Let's quickly sketch our calculation strategy. We will show $A$ is similar to the matrix
\[
A' = \left[ \begin{array}{ccccc}
-(a + b) & a  & & &    \\
b & -(a + b) & a & &  \\
&  &  \ddots  & &        \\
& & b & -(a + b) & 0   \\
& & & b & 0 
\end{array} \right]
\]
and then calculate the eigenvalues of this matrix, which is a much easier task than directly finding the eigenvalues of $A$. We then scale the eigenvalues of $A$ to recover the eigenvalues of $K$.

\subsection*{Transformation of $A$ to $A'$}

To show that $A$ is similar to $A'$, we find a transformation $P$ such that
\[
A' = P^{-1} A P
\]
We assert that the matrix
\[
P = \left[ \begin{array}{ccccc}
1 & 1  & 1& \cdots & 1   \\
 & 1 & 1 &\cdots & 1 \\
&  &  \ddots  & &  \vdots  \\
& &  & 1 & 1   \\
& & &  & 1 
\end{array} \right]
\]
with inverse
\[
P^{-1} = \left[ \begin{array}{ccccc}
1 & -1  & & &    \\
 & 1 & -1 & &  \\
&  &  \ddots  & \ddots &    \\
& &  & 1 & -1   \\
& & &  & 1 
\end{array} \right]
\]
satisfies this requirement. That $A' = P^{-1} A P$ is readily verified via a tedious but simple calculation. Similarly, showing $P^{-1} P = I$ is trivial but tedious. A more elegant method for the verification of these facts surely exists, but has not been given much effort since the brute-force method is quite effective. 

\subsection{The First Eigenvalue of $A'$}

Now, let us find the eigenvalues of $A'$. First, apply Eq. (\ref{continuant}) once to $A' - \lambda I$ to obtain
\[
A' - \lambda I = - \lambda | A'_{N-1} - \lambda I |
\]
where
\[
A'_{N-1} = \left[ \begin{array}{cccc}
-(a + b) & a  & &     \\
b & -(a + b) & a &   \\
&  &  \ddots  &         \\
& & b & -(a + b)   
\end{array} \right]
\]
this shows that the first eigenvalue of $A$ is $\lambda = 0$, and that the rest of the eigenvalues can be found from the roots of the characteristic polynomial of $A'_{N-1} - \lambda I$. Let us find these roots.

\subsection*{Characteristic Polynomial of $A'_{N-1}$}

We will now find the characteristic polynomial of $A'_{N-1} - \lambda I$. For notational ease, set
\[
t = -(a+b) - \lambda
\]
which are the diagonal elements of $A'_{N-1} - \lambda I$.

From Eq. (\ref{continuant}), the polynomial of $A'_{N-1} - \lambda I$ can be clearly seen to be given by the recurrence relation
\[
P_n = t P_{n-1} - P_{n-2}
\]
to solve this, introduce the characteristic equation
\[
x^2 - tx + 1 =0
\]
which has two (complex) roots
\[
x = \frac{t}{2} \pm i \frac{ \sqrt{ 4 - t^2 }}{ 2 }
\]
which, when recast into polar form such that $x = e^{ \pm i \theta}$ can be written
\[
\cos \theta  = \frac{t}{2} \ \ \  \sin \theta  = \frac{ \sqrt{ 4 - t^2 }}{2}
\]
implying
\[
\tan \theta  =  \frac{ \sqrt{ 4 - t^2 }}{t}
\]
The general solution for the recurrence relation is
\[
P_n = \alpha \cos( m \theta ) + \beta \sin( m \theta )
\]
with $m$ the being the size of $A'_{N-1}$, in this case $N-1$. We will retain $m$ for the moment for conciseness.

Now let us find the coefficients $\alpha$ and $\beta$. From the starting conditions of Eq. (\ref{continuant}), we have
\[
P_0 = \alpha = 1
\]
and
\[
P_1 = \alpha \cos( m \theta ) + \beta \sin( m \theta ) = t
\]
substitute our expressions for $\alpha$, $\cos \theta$ and $\sin \theta$ to get
\[
\beta = \frac{ \sqrt{ 4 - t^2 }}{t} = \tan \theta 
\]
such that our final solution is
\begin{equation}\label{poly}
P_n = \cos( m \theta ) + \frac{ \sin( m \theta ) }{ \tan \theta }
\end{equation}
with $\theta$ explicitly given by
\[
\theta = \cos^{-1} \left( - \frac{a + b + \lambda}{ 2 } \right)
\]

\subsection*{Converting the Roots into Eigenvalues of $K$}

Our aim is to find the zeros of $P_n$, which are the eigenvalues of interest. To do this, rearrange Eq. (\ref{poly}) to get
\[
\sin \theta \cos ( m \theta ) + \cos \theta \sin( m \theta ) = 0
\]
which by a common identity is
\[
\sin( \theta + m \theta) = \sin( \theta + (N-1) \theta) = 0
\]
where we have re-substituted $m=N-1$. This is satisfied when $N \theta = j \pi$, with $j$ one of the integers.

These values of $\theta$ our the roots of interest, but we want them in terms of $\lambda$. To convert them, recall that before we had $\cos \theta  = t/2$ and $t = -(a+b) - \lambda$. Combine these with the expression from $\theta$ to obtain
\[
\lambda_A = -(a+b) + 2 \cos \left( \frac{j \pi}{ N }  \right)
\]
where a subscript $A$ indicates these are the eigenvalues of $A'$ (and therefore $A$). In this form, it is clear that we must restrict the values of $j$ to obtain $N-1$ unique eigenvalues. Due to periodicity, there are many choices that will suffice, but the integers $k = 2, 3, ..., N$ will be natural, since then $k$ is an appropriate eigenvalue index (recall we must include $\lambda_1 = 0$ as the first eigenvalue).

To convert the $\lambda_A$ into the eigenvalues of $K$, we need to scale them by $(k_f k_u)^{1/2}$. Substitute for $a$ and $b$, and multiply by this scaling factor to obtain
\[
\lambda = (k_f k_u)^{1/2} \left[ - \sqrt{k_f / k_u } - \sqrt{k_u / k_f } + 2 \cos \left( \frac{k \pi}{ N }  \right) \right]
\]
which reduces to (\ref{spectrum}), our stated solution.

\section*{acknowledgements}
We must thank the very helpful notes Leonardo Volpi has made publicly available on this topic, which are much clearer than any of the published formal treatments. TJ would also like to acknowledge a discussion with Attila Szabo about the conditions for single-exponential behavior, which generated sufficient interest to pursue this problem.

\bibliography{papers2.bib}

\end{document}